# SAMA-IR: Comprehensive Input Refinement Methodology for Optical Networks with Field-Trial Validation

Yihao Zhang
*State Key Laboratory of Advanced Optical Communication Systems and Networks*
*Shanghai Jiao Tong University*
Shanghai, China
yihao.zhang@sjtu.edu.cn

Qizhi Qiu
*State Key Laboratory of Advanced Optical Communication Systems and Networks*
*Shanghai Jiao Tong University*
Shanghai, China
qizhi.qiu@sjtu.edu.cn

Xiaomin Liu
*State Key Laboratory of Advanced Optical Communication Systems and Networks*
*Shanghai Jiao Tong University*
Shanghai, China
xiaomin.liu@sjtu.edu.cn

Jiaping Wu
*State Key Laboratory of Advanced Optical Communication Systems and Networks*
*Shanghai Jiao Tong University*
Shanghai, China
wujiaping@sjtu.edu.cn

Lilin Yi
*State Key Laboratory of Advanced Optical Communication Systems and Networks*
*Shanghai Jiao Tong University*
Shanghai, China
lilinyi@sjtu.edu.cn

Weisheng Hu
*State Key Laboratory of Advanced Optical Communication Systems and Networks*
*Shanghai Jiao Tong University*
Shanghai, China
wshu@sjtu.edu.cn

Qunbi Zhuge*
*State Key Laboratory of Advanced Optical Communication Systems and Networks*
*Shanghai Jiao Tong University*
Shanghai, China
qunbi.zhuge@sjtu.edu.cn

***Abstract*—We propose a novel input refinement methodology incorporating sensitivity analysis and memory-aware weighting for jointly refining numerous diverse inputs. Field trials show ~2.5 dB and ~2.3 dB improvements in Q-factor and power estimation, respectively.**

***Keywords—input refinement, digital twin, field trial***

## I. Introduction

The implementation of digital twins (DTs) plays a pivotal role in building low-margin optical networks to meet ever-increasing capacity demands [1], and accurate physical-layer modeling is vital for building a DT. However, input parameter uncertainties can lead to significant discrepancies between the DT and the real network. To address this challenge, the input refinement (IR) technique has been developed, with extensive research focusing on refining some of the DT inputs [2–4].

However, practical systems pose a more complex challenge for IR, as they typically comprise multiple cascaded components along the link, such as transceivers, fibers, and amplifiers. Each of these components introduces its own set of uncertain input parameters. This multitude of diverse uncertainties coexists and jointly affects the precision of the DT. This necessitates a comprehensive IR approach capable of jointly refining all uncertain inputs. While previous work has made notable advancements [5,6], existing methodologies often employ empirically determined refinement orders for individual inputs, which potentially yield sub-optimal results. Moreover, these approaches may necessitate large data volumes due to the inefficiencies of such heuristic-based schemes [6]. Currently, there is a lack of a systematic methodology to address the IR problem with numerous inputs.

In this paper, we propose SAMA-IR, a comprehensive IR methodology for refining numerous diverse uncertain inputs in optical networks. SAMA-IR employs sensitivity analysis (SA) to prioritize inputs and refine them in descending order of their quantified importance. The process incorporates memory-aware (MA) weighting, ensuring the preservation of previously refined parameters throughout the procedure. This methodology effectively addresses the challenge of joint refinement of numerous inputs. The considered inputs include unknown lumped losses, erbium-doped fiber amplifier (EDFA) characteristics (gain, tilt, and noise figure (NF)), and transceiver impairments. Field-trial experiments demonstrate that SAMA-IR reduces the Q-factor estimation error from -2.52±0.96 dB to -0.03±0.55 dB with only 20 data points. Compared to direct IR, SAMA-IR also achieves a reduction in optical power estimation error by 2.32 dB.

## II. Principle

As shown in Fig. 1(a), SAMA-IR comprises three major steps: 1) The DT is initialized, and the uncertain inputs are identified. 2) SA is conducted to quantify the importance of uncertain inputs. Then the inputs are prioritized based on the resulting importance. 3) The MA-IR is conducted in three stages, where inputs are refined in descending order of importance. Details are introduced as follows.

### A. DT Initialization

First, the DT is constructed based on the field-deployed link configuration depicted in Fig. 1(b). Q-factors are chosen as the quality of transmission (QoT) metric, with their ground-truth values obtainable from the real system. The estimated Q-factor is given by $Q_E = \mathrm{SNR} = 1/(1/\mathrm{GSNR} + 1/\mathrm{SNR}_{\mathrm{TRx}})$, where GSNR is the generalized signal-to-noise ratio (SNR). $\mathrm{SNR_{TRx}}$ denotes the transceiver SNR, acquired from the datasheet and fine-tuned through back-to-back measurements. The amplified spontaneous emission (ASE) noise power is calculated by the purely theoretical model in [7]. The nonlinear interference (NLI) noise power is calculated by GNPy [8], an open-source library base on the Gaussian noise (GN) model. Furthermore, the GSNR-related transceiver penalty is considered in the DT [9].

The considered DT inputs with uncertainties are $X = \{G^{(i)}, T^{(i)}, \mathrm{NF}^{(i)}, \delta^{(j)}, \mathrm{SNR_{TRx}} \mid 1 \le i \le N_{\mathrm{EDFA}}, 1 \le j \le N_\delta\}$, where $G^{(i)}$, $T^{(i)}$, and $\mathrm{NF}^{(i)}$ are the gain, tilt, and NF of the $i$-th EDFA, respectively. $\delta^{(j)}$ is the $j$-th lumped loss along the optical link, as shown in Fig. 1(b). $N_{\mathrm{EDFA}}$ and $N_\delta$ are the number of EDFAs and lumped loss points, respectively.

### B. Sensitivity Analysis

Given the high dimension of $X$, direct optimization of this parameter set is challenging. To address this, SA is employed to assess and prioritize inputs based on their impact on the QoT. The Elementary Effects method is utilized for global SA [10], which provides two key metrics: $\mu^*$, indicating the overall importance of an input, and $\sigma$, representing the input's interactions with other inputs. The importance of the i-th input to the model $F_{\mathrm{DT}}(\cdot)$ is

$$\mu_i^* = \frac{1}{N}\sum_{r=1}^{N_r} \frac{F_{\mathrm{DT}}(x_1,\cdots,x_i+\Delta,\cdots,x_n) - F_{\mathrm{DT}}(x_1,\cdots,x_i,\cdots,x_n)}{\Delta}, \quad (1)$$

where $N_r$ is the number of different samples in the n-dimensional input space, and $\Delta$ is a perturbation.

Fig. 2(a) illustrates the results of the SA, and inputs are categorized into three tiers based on their importance. As shown in Fig. 2(a), tier 1 comprises $G^{(1)} \sim G^{(4)}$, which exhibit high $\mu^*$ and $\sigma$ values, indicating their significant impact on the Q-factor and strong interactions with other inputs. Tier 2 includes $G^{(5)}$, $\delta^{(1)} \sim \delta^{(8)}$, and $\mathrm{SNR_{TRx}}$, which are of moderate importance. The remaining inputs are categorized in tier 3.

### C. Memory-aware IR

Subsequently, a three-stage refinement process is implemented: Stage 1 refines tier 1, Stage 2 addresses tiers 1 and 2, and Stage 3 encompasses all tiers. The IR process for each stage aims at minimizing the loss function

$$\mathcal{L}(Q_E, Q_{\mathrm{meas}}, P_E, P_{\mathrm{meas}}) = \sqrt{\frac{\sum_{i=1}^{N_s}\left(Q_E^{(i)} - Q_{\mathrm{meas}}^{(i)}\right)^2}{N_s} + \lambda_P \sum_{i=1}^{2N_{\mathrm{EDFA}}}\left(P_E^{(i)} - P_{\mathrm{meas}}^{(i)}\right)^2 + \lambda_{\mathrm{MA}} \sum_{i=1}^{m}\left[\mu_i^*\left(X_i - X_i^{\mathrm{lastopt}}\right)\right]^2}, \quad (2)$$

where $Q_{\mathrm{meas}}$ is the measured Q-factor. $P_E$ and $P_{\mathrm{meas}}$ denote the estimated and measured total signal power, respectively, at the input and output of EDFAs. The first term in Eq. (2) is the root mean square error (RMSE) of Q-factors, which is the primary metric for evaluating the precision of DT. The second term quantifies the optical power estimation error, penalized by a regularization factor $\lambda_P$. The third term is introduced after Stage 1, designated as the MA term. It imposes a weighted penalty on the deviation of input $X_i$ from its previous stage's optimal value (if has been refined), denoted as $X_i^{\mathrm{lastopt}}$. The weighting factor $\mu_i^*$ represents the importance of the input $X_i$. By incorporating this term from Stage 2, the optimization process maintains a "memory" of previous solutions.

$\lambda_P$ is initially set to a relatively high value in Stage 1, given that tier 1 parameters are primarily related to signal power. As the stages progress, $\lambda_P$ is gradually reduced to a minimal value, effectively mitigating the influence of power measurement errors on the IR results. Moreover, the MA term ensures that previously learned power-related parameters are retained throughout the optimization process. The Bayesian optimization (BO) is utilized to solve the optimization problem and determine the optimal $X$, owing to its high efficiency in exploring high-dimensional spaces.

## III. FIELD-TRIAL EXPERIMENT SETUP

Fig. 1(b) illustrates the field-deployed testbed utilizing commercial equipment spanning from Shanghai through Jiaxing to Hangzhou in China. The configuration comprises a 440-km loop link incorporating four spans of G.652.D fibers and six EDFAs. A transmission scenario with 30 wavelengths in a 75-GHz fixed grid is emulated. Six transponders operating at 63.9 GBaud and 200 Gbps are employed at the 3rd, 8th, 13rd, 18th, 23rd, and 28th channels. The other channels are filled with dummy signals. A software-defined network (SDN) controller is implemented. Control commands and millisecond-level telemetry data are transmitted via the NETCONF protocol using

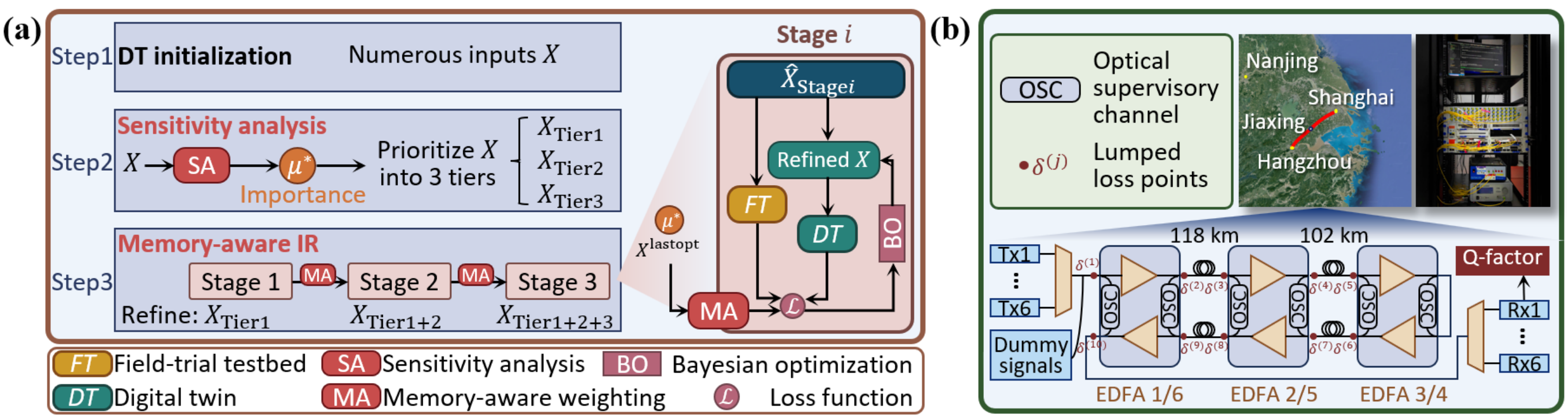

Fig. 1. **(a)** The flowchart of the proposed SAMA-IR scheme. **(b)** The field-trial testbed.

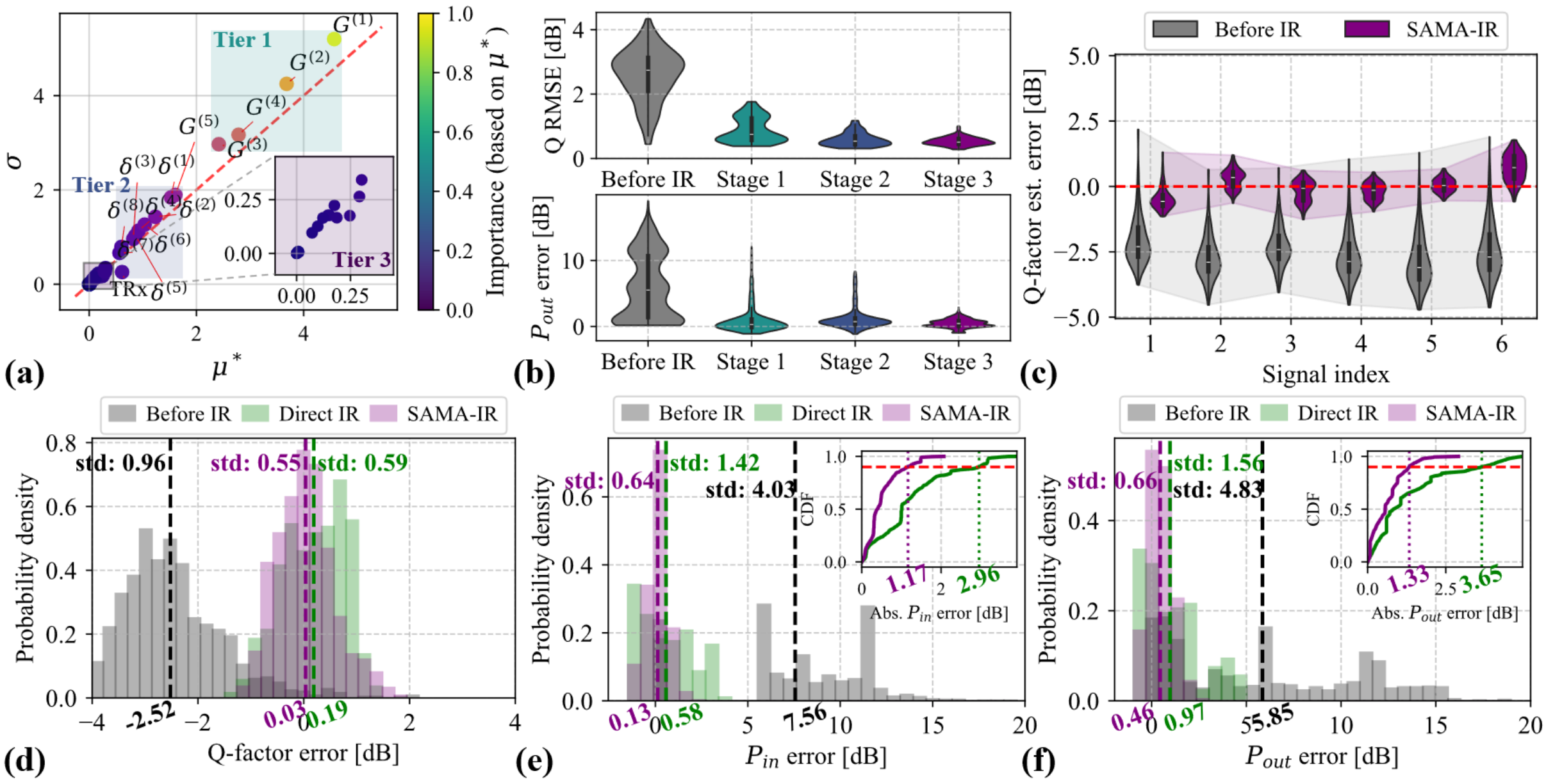

Fig. 2. **(a)** The importance of all inputs given by the Elementary Effects method. **(b)** The RMSEs of Q-factors and the errors of $P_{\mathrm{out}}$ at the end of 3 stages. **(c)** The Q-factor estimation error distribution of all 6 signals. **(d-f)** The error distribution of **(d)** Q-factor, **(e)** $P_{\mathrm{in}}$, and **(f)** $P_{\mathrm{out}}$, estimated by the DT before IR, after direct IR, and after SAMA-IR.

YANG models, which are transmitted through the optical supervisory channel (OSC).

To refine the input parameters, 20 pieces of data are collected with the gains and tilts of all EDFAs randomly set. Another 180 data samples are independently collected for testing. $\lambda_P$ is set to 1.0, 0.2, and 0.01 for stage 1, 2, and 3, respectively. $\lambda_{\mathrm{MA}}$ is set to 0.05. Two baseline schemes are also evaluated: 1) *Before IR*, where all inputs are acquired from measurements or datasheets; 2) *Direct IR*, where all IR inputs are simultaneously optimized. To ensure the fairness, the direct IR also employs BO as the optimization strategy.

## IV. Results

Fig. 2(b) shows the Q-factor and signal power modeling results on the test set using refined input parameters after each of the three stages. Notably, after Stage 1, the accuracy of power modeling shows a greater improvement compared to that of Q-factor modeling due to the large $\lambda_P$ employed in Stage 1. Following Stage 2 and 3, the accuracy of Q-factor estimation is significantly improved. Moreover, the accuracy of power estimation does not deteriorate in the final stage, owing to the implementation of MA. Fig. 2(c) depicts the Q-factor estimation error after Stage 3, demonstrating that SAMA-IR effectively minimizes the error distributions to the vicinity of zero.

Moreover, Fig. 2(d) shows the Q-factor estimation error distributions of the DT before IR, after SAMA-IR, and after direct IR. Results illustrate that the SAMA-IR significantly reduces the mean error from -2.52 dB to 0.03 dB and decreases the standard deviation (std) from 0.96 dB to 0.55 dB, whereas the direct IR method shows slightly inferior performance. However, for power estimation, Fig. 2(e) and 2(f) show that SAMA-IR significantly outperforms the direct IR method, greatly reducing both mean errors and standard deviations. The insets in Fig. 2(e) and 2(f) illustrate the cumulative distribution function (CDF) of the absolute power estimation error. At the 90th percentile of the CDF, SAMA-IR reduces the error by 1.79 dB (from 2.96 dB to 1.17 dB) compared to the direct IR method for EDFA input power, and by 2.32 dB (from 3.65 dB to 1.33 dB) for EDFA output power. These results indicate that SAMA-IR not only achieves higher QoT estimation accuracy, but also shows better physical consistency with the practical system.

## V. Conclusions

An IR methodology combining SA with staged, memory-aware IR to refine numerous diverse uncertain inputs in optical networks is proposed. Field trials show that SAMA-IR significantly reduces the Q-factor estimation error from -2.52±0.96dB to -0.03±0.55dB and enhances optical power estimation accuracy by up to 2.32 dB.

## Acknowledgment

This work was supported by Shanghai Pilot Program for Basic Research - Shanghai Jiao Tong University (21TQ1400213) and National Natural Science Foundation of China (62175145)